\documentclass[aps,prl,preprint,superscriptaddress,showpacs]{revtex4}
\usepackage{graphicx}
\usepackage{bm}
\usepackage{amsmath}
\usepackage{amssymb}
\usepackage{hyperref}

\begin{document}

\title{Deformation of doubly-clamped single-walled carbon nanotubes in an electrostatic field}

\author{Zhao Wang}
\email{zhao.wang@empa.ch,wzzhao@yahoo.fr}
\affiliation{EMPA (Swiss Federal Laboratories for Materials Testing and Research), Feuerwerkerstrasse\,39, CH-3602\,Thun, Switzerland}
\affiliation{Institute UTINAM, Universit\'{e} de Franche-Comt\'{e}, F-25000 Besan\c{c}on, France.}

\author{Laetitia Philippe}
\affiliation{EMPA (Swiss Federal Laboratories for Materials Testing and Research), Feuerwerkerstrasse\,39, CH-3602\,Thun, Switzerland}

\begin{abstract}
In this letter, we demonstrate a strong dependence of the electrostatic deformation of doubly-clamped single-walled carbon nanotubes on both the field strength and the tube length, using molecular simulations. Metallic nanotubes are found to be more sensitive to an electric field than semiconducting ones of the same size. For a given electric field, the induced deformation increases with tube length but decreases with tube radius. Furthermore, it is found that nanotubes can be more efficiently bent in a center-oriented transverse electric field.
\end{abstract}

\pacs{85.35.Kt, 32.10.Dk, 31.15.Qg}


\date{\today}

\maketitle

Carbon nanotubes (CNTs) can be used in nanoelectromechanical systems (NEMS) for uses in sensing, actuation, vibration, and laboratory-on-a-chip applications \cite{Anantram2006}. Particularly, doubly-clamped  (suspended) structures of CNTs have been reported to be used as key components in a number of nanodevices \cite{Javey2002,Javey2003,Keren2003,Sapmaz2003,Sazonova2004,Jonsson2005}. In most of these recent NEMS, nanotubes are usually suspended between two electrodes with an applied voltage. However, as to be demonstrated in this paper, even without applying electric current and excess charges, doubly-clamped CNTs can still be significantly bent in a static transversal electric field, as a result of electric polarization.

Cantilevered CNTs have long been known to deform in an electric field \cite{Poncharal1999,Wei2001a,Wang2007a}. The mechanism of this deformation relies on the fact that CNTs can be bent by a bending moment induced by interactions between the electric field and the molecular dipoles due to electric polarization. This property has been exploited in the design of nanorelays \cite{Kinaret2003,lee-04}. Compared with cantilevered CNTs, doubly-clamped ones can have many advantages in electronic devices \cite{Dai2002} (e.g. they can be integrated in nanodevices with a well-defined spatial structure and resonance frequency). This letter investigates the deflection of neutral single-walled CNTs (SWCNTs) by an electrostatic field. Two semiclassical theories have been combined for characterizing chemical potential and electrostatic interactions in both metallic and semiconducting CNTs. Considering the general correlation between the conductivity and polarizabilities of CNTs, metallic CNTs are supposed to be more sensitive to electric fields than semiconducting ones are, and hence can be expected to play a more important role in NEMS.

When an electric field is applied to a metallic CNT, an induced dipole can be created at each atomic site with a quantity of free charge, by shifting the electrons and the nuclei. In our calculation, each atom is therefore modeled as an interacting polarizable point with a free electric charge, while the chemical bonds are described by using a many-body potential function. Motionless equilibrium positions of carbon atoms in an electric field are computed by minimizing the total potential energy of systems $U^{tot}$, which is the sum of two terms: $U^{tot}=U^{elec}+U^{p}$, where $U^{p}$ is the interatomic potential due to the C-C chemical bonds in absence of an external field, including the long-range interactions. $U^{elec}$ stands for the electrostatic energy from the interaction between charges, dipoles and external fields. $U^{elec}$ is calculated using a Gaussian-regularized charge-dipole \cite{Wang2007a} and a dipole-only \cite{zhaowang-06-01} model for metallic and semiconducting CNTs, respectively. $U^{p}$ is computed using an adaptive interatomic reactive empirical bond order (AIREBO) potential \cite{Stuart2000}, which is an extension of the REBO potential \cite{Brenner1990b} which includes a Lennard-Jones potential. Compared to first-principle and semi-empirical methods, an important feature of our combined models is their ability to deal with large systems (up to $5000$ atoms). This is particularly important for studying geometric effects on the electrostatic bending of CNTs, since the axial periodic condition can hardly be applied in this issue.

The strength of electric field $E$ used in this work is in the order of V/nm, because the tubes used in our calculation are too short ($18 \backsim 27$ nm) to be bent in a weak electric field. As will be shown, the required field strength decreases with increasing tube length $L$ for a given deformation. Considering that the CNTs used in experiments are usually of $\mu$m-length, field emission effects \cite{Jo2003} and conductance switching due to strong transversal fields \cite{Son2005} are both neglected. These field strengths are not large enough to cause a significant change in the chemical bond strength \cite{Guo2003a}. Moreover, it can expected that an armchair CNT can exhibit more deformation by an electric field than a metallic zigzag tube of the same size, since previous studies concluded that bending deformation would have a negligible effect on the electronic transport properties of armchair CNTs \cite{Rochefort1999,Nardelli1999}, while the quantum conductance of zigzag CNTs can significantly decrease under large bending deformations \cite{Maiti2002}. In this work, the tubes are assumed to be suspended between two electrically insulated supports. 

\begin{figure}[ht]
\centerline{\includegraphics[width=14cm]{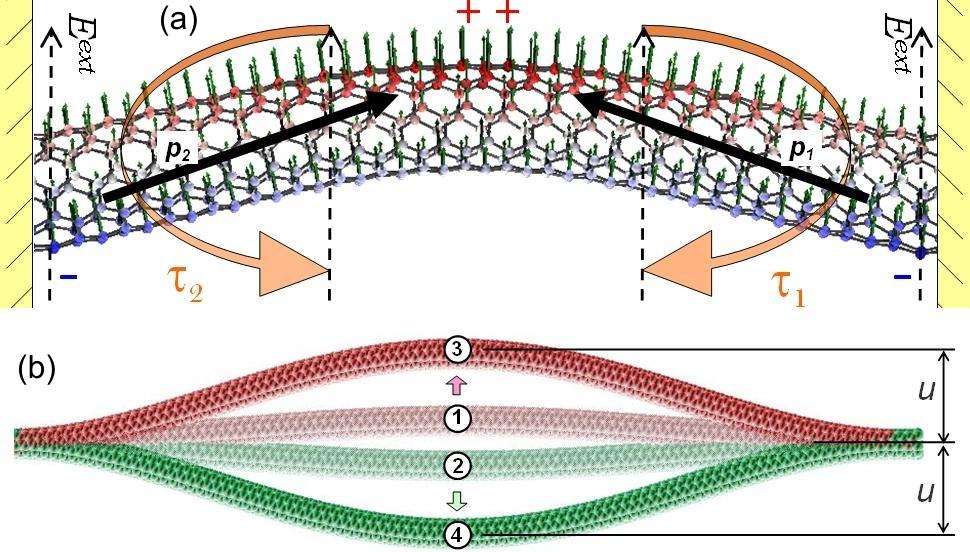}}
\caption{\label{fig:schemadef}
(Color online) Schematics of electrostatic deformation of a SWCNT. (a) Two arrows $\bm{p}_{1}$ and $\bm{p}_{2}$ stand for two induced molecular dipoles.  $\bm{E}^{ext}$ is represented by the arrows with dashed lines. $\tau_{1}$ and $\tau_{2}$ stand for the induced bending moments acting on the tube. The color scale of atoms is proportional to the density of induced charges. The vectors stand for the induced atomic dipoles. (b) Two possible deflection directions in bending (upward or downward).}
\end{figure}

The mechanism of electrostatic bending of a SWCNT is depicted in Fig. \ref{fig:schemadef} (a). When a CNT (initially electrically neutral) is submitted to a transversal electric field $\bm{E}^{ext}$, the field tends to shift negative and positive charges in opposite directions (in this figure we can see that positive charges move to the top and negative ones move to the bottom of the tube), two induced molecular dipoles ($\bm{p}_{1}$ and $\bm{p}_{2}$) are hence created in the tube. The tube is then bent by two bending moments ($\tau_{1}$ and $\tau_{2}$), which is induced by the interaction between the molecular dipoles and $\bm{E}^{ext}$ ($\tau = \bm{p}^{m} \otimes \bm{E}^{ext}$). 

Molecular dynamics (MD) simulations were performed using the AIREBO potential, for generating arbitrary initial configurations of CNTs at room temperature. The energy of these MD-generated configurations is minimized in order to calculate their equilibrium positions in an electric field. MD simulations showed that CNTs slightly oscillate around their initial positions due to thermal fluctuations, which have been observed in experiments \cite{Treacy1996,Babic2003}. We found that the orientation of electrostatic deformation of CNTs strongly depends on their initial position at the moment when $\bm{E}^{ext}$ is applied, e.g., as shown in Fig. \ref{fig:schemadef} (b), if $\bm{E}^{ext}$ is applied when the tube is at position $1$, its equilibrium state in electric field will be at position $3$; conversely if the tube is initially in position $2$, it will be bent to position $4$.  These results also imply that no change of the deformation direction will take place if the field direction is totally reversed, due to the system symmetry. For measuring the amplitude of deflection, we define $u$ as the displacement of the center of the tube middle from its initial position. For a small electric field, the bending of the CNT is elastically reversible, i.e., the tube will come back and oscillate around its initial position once $\bm{E}^{ext}$ is removed \cite{Sazonova2004,Wei2001}. In our MD simulations, the average frequency of thermal oscillation (without electric field) of a (5, 5) tube (length $L\approx20.0$ nm) is found to be $60 \pm 10$ GHz for the several first harmonics. This value is comparable to oscillation frequency of doubly-clamped CNTs recently reported in Ref. \cite{Li2003}.   

\begin{figure}[ht]
\centerline{\includegraphics[width=12cm]{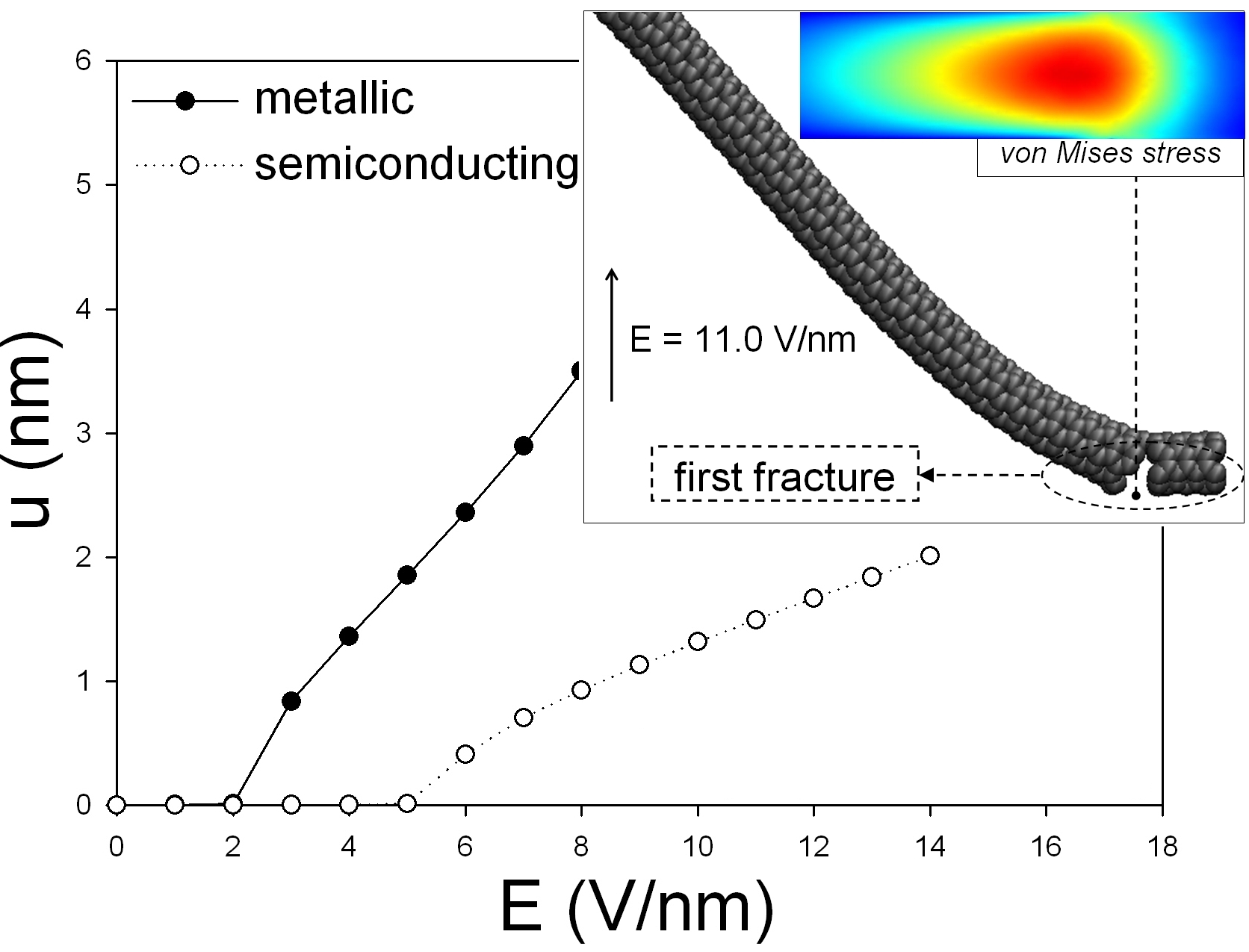}}
\caption{\label{fig:strength_break}
(Color online) Electrostatic deformation $u$ versus field strengths $E=\left| \bm{E}^{ext} \right|$ for two SWCNTs: a metallic (4, 4) ($L \approx 24.5$ nm) and a semiconducting (5, 3) with almost the same length and radius. The tube (4, 4) begins to break down when $E > 11.0$ V/nm, then failure occurs very soon thereafter ($u_{max} > 5.1$ nm). The inset shows the position of fracture and the distribution of von-Mises stress at the bottom (before the fracture occurs).}
\end{figure}

We plot in Fig. \ref{fig:strength_break} computed values of $u$ versus external field strength $E$ for both a (4, 4) and a (5, 3) CNTs. It can be seen that the metallic tube is clearly more sensitive to $E^{ext}$ than the semiconducting one due to their different polarizabilities \cite{Joselevich2002}. When $u \approx 5.5$ nm, it is observed that the (4, 4) tube begins to fracture near the fixed end, in which the maximal tensile strain is localized (inset in Fig. \ref{fig:strength_break}). We can also see that there is no deformation when $E$ remains small, since in such cases the moment of electric force are not large enough to exceed the mechanical resistance barrier of the tube. These results also imply that semiconducting tubes can sustain higher field strengths.

\begin{figure}[ht]
\centerline{\includegraphics[width=11cm]{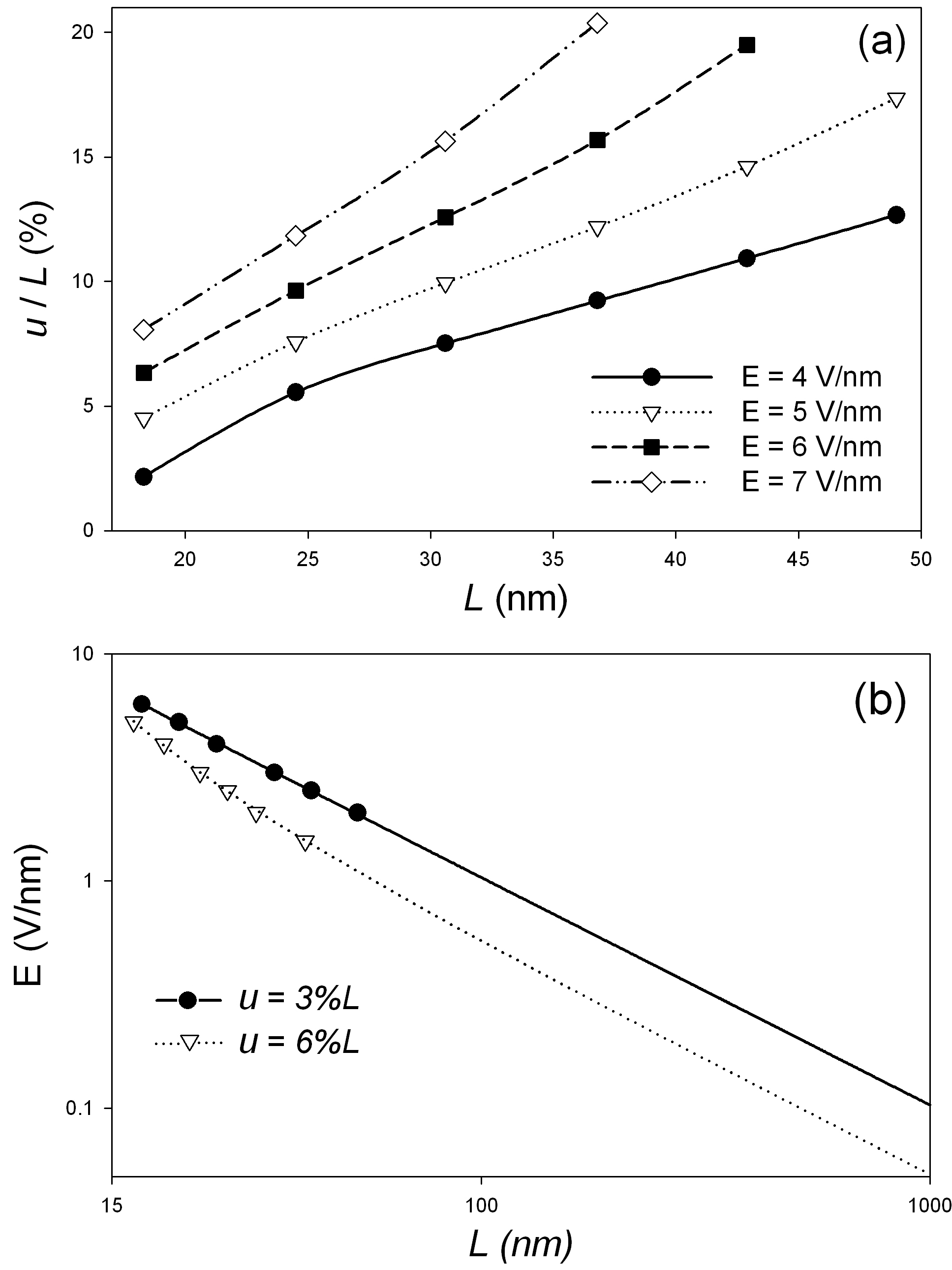}}
\caption{\label{fig:lengthdep}
(a) $u/L$ versus the tube length $L$ for (4, 4) SWCNTs, in 4 different transverse $E^{ext}$  (which is perpendicular to the initial axis of the tubes). (b) $E$ versus $L$ on a logarithmic scale for two given $u$. The symbols present the calculated points and the lines stand for the extrapolation curves.}
\end{figure}

The CNTs used in experiments are usually microns in length. However, their field-induced deformations can hardly be addressed by the calculations using atomic models in a direct way. Hence, it is necessary to address the influence of the tube length $L$. In Fig. \ref{fig:lengthdep} (a), it can be seen that the relative displacement $u/L$ increases with $L$, i.e., a weaker electric field is required for a given $u$ in a longer tube. This increased displacement depends on the field strength. An extrapolation was done from the data $E$ versus $L$ for tube lengths up to $1$ $\mu$m, in order to make experimental verification possible (Fig. \ref{fig:lengthdep} (b)). It was found that $\ln(E)$ decreases almost linearly with $\ln(L)$ for two given $u$. This is related to the fact that the electric capacitance per unit length of a metallic cylinder is roughly proportional to $\ln(L)$. Thus, we can conclude that, for the $\mu$m-length tubes used in experiments, the required field strength is on the order of V$/\mu$m.

\begin{figure}[ht]
\centerline{\includegraphics[width=12cm]{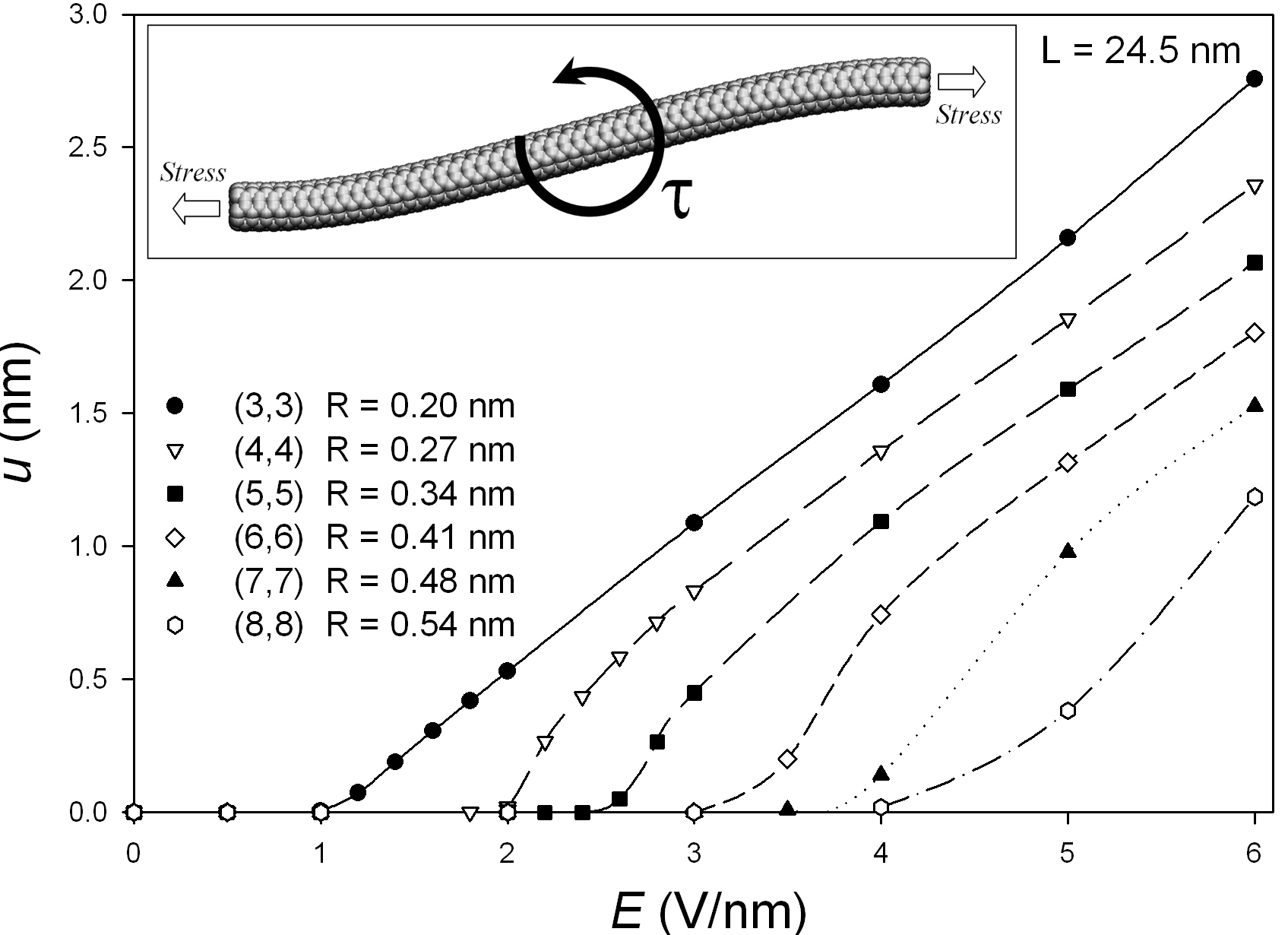}}
\caption{\label{fig:radiuseb}
$u$ versus $E$ for 6 individual armchair SWCNTs ($L \approx 24.5$ nm) with different radii. The direction of $E^{ext}$ is perpendicular to the initial axis of tubes. Inset shows the force balance between an electric moment and internal stresses in a half of a deformed tube.}
\end{figure}

To explore further geometry effects, we plot $u$ versus $E$ for the tubes of different radii $R$ in Fig. \ref{fig:radiuseb}. It is found that small tubes are more sensitive to $E^{ext}$ than large ones. This looks abnormal since it is known that the polarizability of a CNT increases with its radius. However, from a mechanical point of view, the tube becomes more difficult to be bent due to the increase of its area moment of inertia ($I=0.5m(R^{2}_{int }+R^{2}_{out})$ for a thick-walled cylindrical tube, where $m$ is the mass). From our results it is clear that the later effect plays a more important role. The ratio $u/E$ is roughly the same for all of these tubes in case of large deformation ($u>1$ nm). 

In Figs. \ref{fig:strength_break} and \ref{fig:radiuseb}, we can see a threshold field of a few V/nm for each $u$ versus $E$ curve before one obtains a nonzero deformation. The existence of this threshold field relies on the fact that the electric bending moments $\tau$ remains weaker than the tube resistance when $E$ is small. With increasing $E$, one reaches the limit of the threshold field until a value of $u$ allows a moment balance in the nanotube (inset in Fig. \ref{fig:radiuseb}). It is found that the threshold field increases with $R$ but decreases with $L$.  

\begin{figure}[ht]
\centerline{\includegraphics[width=12cm]{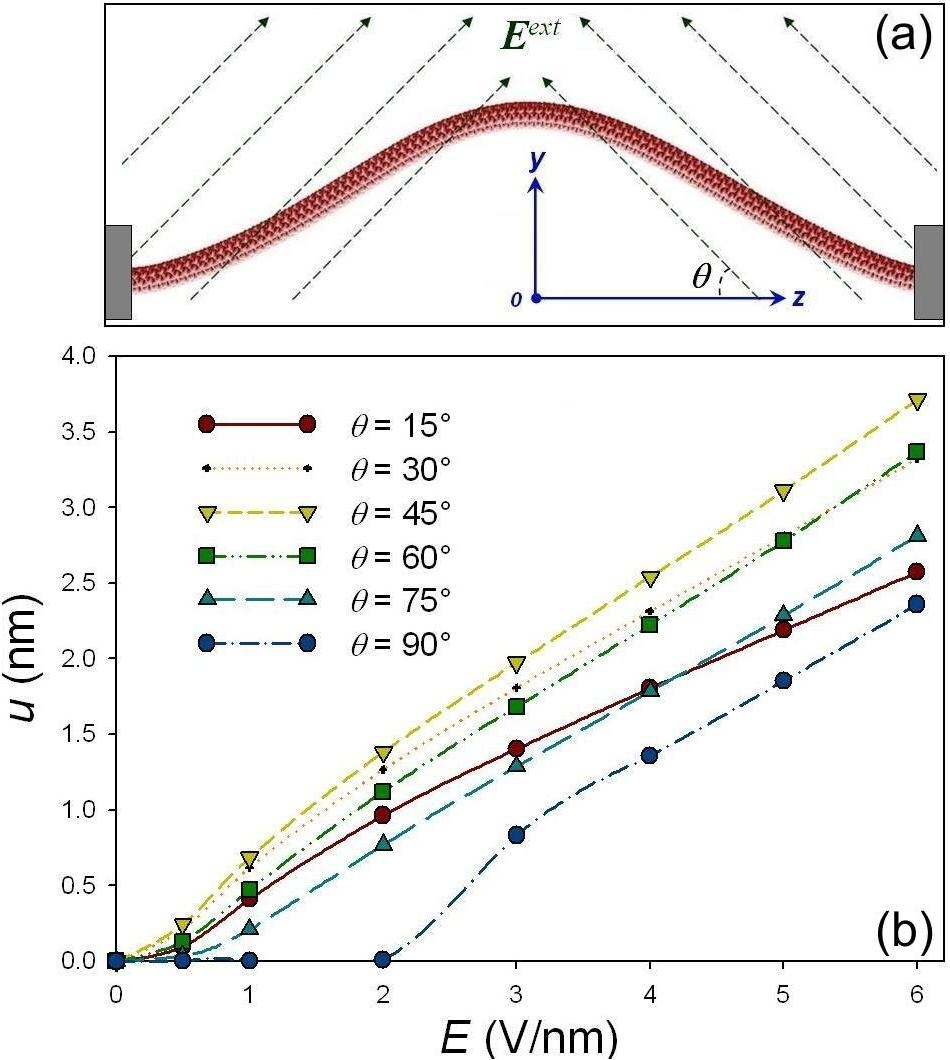}}
\caption{\label{fig:directioneb}
(Color online) (a): Schematic (in \textit{y-z} plane) of the deformation of a CNT subjected to an $\nearrow\,\nwarrow$-like electric field $E^{ext}$ (dashed lines). The field angle $\theta$ is defined as the angle between the field lines and the axis $z$. The field vector $\bm{E}=(E_{x},\,\,E_{y},\,\,E_{z}) = (0,\,\,E\sin(\theta),\,\,+E\cos(\theta))$ when $z<0$ and $\bm{E}=(0,\,\,E\sin(\theta),\,\,-E\cos(\theta))$ when $z>0$. (b): $u$ versus $E$ for a (4, 4) SWCNT ($L\approx24.5$ nm) in $E^{ext}$ with different $\theta$. The simulation data are represented by the symbols with eye-guide lines.}
\end{figure}

In this study, the applied external field is assumed to be uniform as a common theoretical simplification. However, in experimental situations, the external fields are usually not homogeneous and strongly depend on the experimental set-up. By changing the field direction, we found that CNTs can be more efficiently bent in a center-oriented ($\nearrow\,\nwarrow$-like) electric field (see Fig. \ref{fig:directioneb} (a)). Results of the deformation of a SWCNT by this field is plotted in Fig.\ref{fig:directioneb} (b). We can see that values of $u$ roughly follow a linear relationship with $E$ when $u>1$ nm. It was also found that, when $\theta=\pi/4$, the tube can be most efficiently bent and its $u$ versus $E$ curve has straightest shape. 

In summary, deformation of doubly-clamped SWCNTs in an electrostatic field has been simulated using a charge-dipole polarization model combined with an empirical potential. The interplay between the mechanical resistance and electric polarization of CNTs was investigated for both metallic and semiconducting tubes. Metallic CNTs are found to be more sensitive to an electric field than semiconducting ones. This study reveals that the field-induced deflection increases with tube length while it decreases with tube radius. It was also found that CNTs can be more efficiently bent in a center-oriented electric field. From a theoretical point of view, graphene nanoribbons and other metallic nanowires/tubes should have similar properties, depending on their dielectric constants. This electrostatic deformation of doubly-clamped nanostructures can be expected to open a path to designs of novel nanodevices.

We gratefully thank S. J. Stuart for his help to the implementation of our computational code. M. William, A. Mayer, R. Langlet, M. Devel and W. Ren are acknowledged for useful discussions.

\end{document}